\documentclass[12pt,a4j]{article}
\usepackage[dvips]{graphicx}
\usepackage{enumerate}
\usepackage{amsmath}

\begin{document}\title{Coupled Dispersionless and Generalized Heisenberg Ferromagnet Equations with Self-Consistent Sources:  Geometry and  Equivalence}
\author{Guldana Bekova\footnote{Email: bekovaguldana@gmail.com}, \, Gulgassyl Nugmanova\footnote{Email: nugmanovagn@gmail.com}, \,   Gaukhar Shaikhova\footnote{Email:  g.shaikhova@gmail.com}, \\   Kuralay Yesmakhanova\footnote{Email: kryesmakhanova@gmail.com} \,    and Ratbay Myrzakulov\footnote{Email: rmyrzakulov@gmail.com}\\
\textsl{Eurasian International Center for Theoretical Physics and} \\ { Department of General \& Theoretical Physics, } \\ Eurasian  National University,
Astana, 010008, Kazakhstan
}
\maketitle
\begin{abstract}
We propose a  new integrable  coupled dispersionless  equation with self-consistent sources (CDESCS).  We obtain the Lax pair  and the equivalent generalized Heisenberg ferromagnet equation  (GHFE), 
demonstrating its  integrability.  Specifically, we explore the geometry  of these equations. 
Last, we consider the relation between the motion of curves/surfaces  and the CDESCS and the GHFE.
\end{abstract}



\section{Introduction} 
The study of integrable systems  or solvable nonlinear  differential equations (NDE)
is an active area of research since the discovery of the inverse scattering method. These equations  are in a sense universal since they show up in many areas of physics and  mathematics.  As integrable
systems we understand those which have infinite hierarchy of symmetries and conservation laws.  In the theory of solvable  nonlinear differential equations,  one of the most important issues is a systematic method for construction of integrable systems. For integrable systems there exist several parallel schemes of construction.   

Besides the integrable NDEs, there is another important class of integrable
partial differential equations: the so-called ”integrable equations of hydrodynamic type”, often called ”dispersionless equations” \cite{SA1}-\cite{Pavlov}. In some cases, these equations  are the dispersionless (or semiclassical) limits of integrable soliton systems, or by construction, such as the system of hydrodynamic type.  They often arise in various
problems of physics and mathematics, for this reason, they  are intensively studied in the recent
years.    Some of these equations are integrable in the Hamiltonian sense. Integrable dispersionless equations  are equivalent to the
commutation condition of vector fields Lax pairs, so that  they
can be in an arbitrary number of dimensions. For example,  in   \cite{SA1}-\cite{SA3},  a new systematic method was introduced for construction of dispersionless systems in 3+1 dimensions  using nonisospectral Lax pairs that involve
contact vector fields.

In this paper we deal with the so-called  Konno-Oono equation (KOE) \cite{Konno1}
\begin{eqnarray}
q_{xt}-\rho q&=&0, \label{1} \\
\rho_{x}+0.5(q^{2})_{t}&=&0.  \label{2}
\end{eqnarray}
The KOE (\ref{1})  is also  known as   the coupled dispersionless equation (CDE). Later,  Konno and Kakuhata proposed a generalization of this equation in the following form  \cite{Konno2}
\begin{eqnarray}
q_{xt}-\rho q&=&0, \label{3} \\
r_{xt}-\rho r&=&0, \label{4} \\
\rho_{x}+0.5(rq)_{t}&=&0  \label{5}
\end{eqnarray}
which is  called  the generalized Konno-Oono equation (GKOE) or the generalized CDE (GCDE). Its  Lax representation (LR) can be expressed as
\begin{eqnarray}
\Psi_{x}=U_{1}\Psi,\quad 
\Phi_{t}=V_{1}\Psi,\label{6}
\end{eqnarray}
where
\begin{eqnarray}
V_{1}=-i\lambda \left(
\begin{array}{cc}
\rho& q_{t} \\
r_{t} & -\rho
\end{array}
\right), \quad U_{1}=\left(
\begin{array}{cc}
\frac{i}{4\lambda} & -0.5q \\
0.5r& -\frac{i}{4\lambda}
\end{array}
\right). \label{7}
\end{eqnarray}
It were pointed that  the KOE (\ref{1}) is gauge equivalent to the sine-Gordon equation, whereas its  complex version that corresponds to the GKOE (\ref{3})-(\ref{5}) with the reduction $r=\bar{q}$  is gauge equivalent to the Pohlmeyer-Lund-Regge equation \cite{Kotlyarov}. It is interesting to note that there exist  other gauge and geometrical equivalent equations to  the GKOE (\ref{3})-(\ref{5}) and to its reductions. These equivalent equations are some kind of generalizations of the famous  Heisenberg ferromagnet equations (HFE) 
\cite{l77}-\cite{Takhtajan}
\begin{eqnarray}
iA_{t}=\frac{1}{2}[A,A_{xx}]\label{8} 
\end{eqnarray} 
and which are the subject of this paper. 

The integrable systems  with self-consistent sources (ISSCS) have attracted some attention (see, for example, [1]-[14]). The ISSCS can be solved by the inverse scattering transform (IST) method,  and
N-soliton solutions of some ISSCSs were obtained [1]-[15].

The present paper is organized as follows. In Sec. 2, we present the M-XIV  equation in the matrix and the vector form. 
 Specifically, we employ a Lax representation of this equation.
 In Sec. 3, we demonstrate  the geometric  representation of the M-XIV equation. In particular, we show that the M-XIV equation and the coupled dispersionless equation with self-consistent equation  are closely related to each other.  In the following Sec. 4 we establish the gauge equivalence between the M-XIV and   M-XXXII equations.
In Sec. 5, we analyze several reductions of the  M-XXXII equation in detail and  present their  Lax representations.  The scale transformations of the  M-XXXII equation is given in Sec. 6.  Integrable surface induced by the M-XIV  equation is presented in Sec. 7. The corresponding surface equation is the M-XXXI equation.  
 Section 8 is devoted to concluding remarks.

\section{Myrzakulov-XIV equation}
In this section, we present  the different forms of the  Myrzakulov-XIV (M-XIV) equation as well as its Lax representation, reduction and simplest conservation law.

\subsection{Equation}
\subsubsection{The matrix M-XIV equation}
Consider the following M-XIV equation
\begin{eqnarray}
iA_{t}&=&i(fA)_{x}+\frac{1}{4\alpha}[A,A_{t}]_{x}+\frac{1}{\alpha}[A,W], \label{9} \\
W_{x}&=&i(\alpha-\omega)[A,W] \label{10} 
\end{eqnarray}
or
\begin{eqnarray}
iA_{t}&=&ifA_{x}+\frac{1}{4\alpha}[A,A_{tx}]+\frac{1}{\alpha}[A,W], \label{11} \\
f_{x}&=&\frac{1}{4i\alpha}tr(A[A_{t},A_{x}]), \label{12} \\
W_{x}&=&i(\alpha-\omega)[A,W]. \label{13} 
\end{eqnarray}
Here 
\begin{eqnarray}
A&=&\left(
\begin{array}{cc}
A_{3} & \sigma A^{-} \\
A^{+} & -A_{3}
\end{array}
\right), \quad A^{2}=I, \quad A^{\pm}=A_{1}\pm i A_{2}, \quad A_{1}^{2}+A^{2}_{2}+A^{2}_{3}=1,  \label{14}\\
W&=&\left(
\begin{array}{cc}
W_{3} & \sigma W^{-} \\
W^{+} & -W_{3}
\end{array}
\right), \quad W^{2}=I, \quad W^{\pm}=W_{1}\pm i W_{2}, \quad W_{1}^{2}+W^{2}_{2}+W^{2}_{3}=1. \label{15}
\end{eqnarray}
\subsubsection{The vector  M-XIV equation}
In the vector form the M-XIV equation looks like
\begin{eqnarray}
{\bf A}_{t}&=&(f{\bf A})_{x}+\frac{1}{2\alpha}({\bf A} \wedge {\bf A}_{t})_{x}+\frac{2}{\alpha}{\bf A} \wedge {\bf W}, \label{16}\\
{\bf W}_{x}&=&2(\omega-\alpha){\bf A} \wedge {\bf W} \label{17}
\end{eqnarray}
or
\begin{eqnarray}
{\bf A}_{t}&=&f{\bf A}_{x}+\frac{1}{2\alpha}{\bf A} \wedge {\bf A}_{xt}+\frac{2}{\alpha}{\bf A} \wedge {\bf W}, \label{18}\\
f_{x}&=&\frac{1}{2\alpha}{\bf A}\cdot ({\bf A}_{t} \wedge {\bf A}_{x}), \label{19}\\
{\bf W}_{x}&=&2(\omega-\alpha){\bf A} \wedge {\bf W}, \label{20}
\end{eqnarray}
where
\begin{eqnarray}
{\bf A}&=&(A_{1}, A_{2}, A_{3}),\quad {\bf A}^{2}=1, \quad  f=\frac{1}{\alpha^{2}}a+\frac{1}{\alpha(\alpha-\omega)}\eta, \label{21}\\
{\bf W}&=&(W_{1}, W_{2}, W_{3}), \quad 
W_{1}^{2}+W_{2}^{2}+W_{3}^{2}=const=1, \quad {\bf A}\cdot {\bf W}_{x}=0. \label{22}
\end{eqnarray}
\subsubsection{The M-XIV equation as the equation with self-consistent sources}
We can   rewrite the M-XIV equation as the equation with self-consistent sources. Let us we introduce  the following representation for the components of the matrix function $W$:
\begin{eqnarray}
W_{1}&=&\phi_{1}\bar{\phi}_{2}+\bar{\phi}_{1}\phi_{2}, \\\label{23}
W_{2}&=&i(\phi_{1}\bar{\phi}_{2}-\bar{\phi}_{1}\phi_{2}), \\ \label{24}
W_{3}&=&|\phi_{1}|^{2}-|\phi_{2}|^{2} \label{25}
\end{eqnarray}
or
\begin{eqnarray}
W^{+}&=&2\bar{\phi}_{1}\phi_{2}, \\\label{26}
W^{-}&=&2\phi_{1}\bar{\phi}_{2}, \\\label{27}
W_{3}&=&|\phi_{1}|^{2}-|\phi_{2}|^{2},\label{28}
\end{eqnarray}
so that   the matrix $W$ is given by
\begin{eqnarray}
W=\left(\begin{array}{cc}
W_{3} &  W^{-} \\
W^{+} & -W_{3}
\end{array}\right)=\left(
\begin{array}{cc}
|\phi_{1}|^{2}-|\phi_{2}|^{2} & 2\phi_{1}\bar{\phi}_{2}\\
2\bar{\phi}_{1}\phi_{2} & |\phi_{2}|^{2}-|\phi_{1}|^{2}
\end{array}
\right). \label{29}
\end{eqnarray}
Here  $\phi_{j}$ obey the following set of equations
\begin{eqnarray}
\phi_{1x}&=&-i\zeta(A_{3}\phi_{1}+A^{-}\phi_{2}), \\ \label{30}
\phi_{2x}&=&-i\zeta(A^{+}\phi_{1}-A_{3}\phi_{2}), \label{31}
\end{eqnarray}
where $\zeta=\omega-\alpha$. Thus  the M-XIV equation (as the equation with self-consistent sources) can be written as 
\begin{eqnarray}
iA_{t}&=&ifA_{x}+\frac{1}{4\alpha}[A,A_{tx}]+F,  \\ \label{32}
f_{x}&=&\frac{1}{4i\alpha}tr(A[A_{t},A_{x}]),  \\ \label{33}
\phi_{1x}&=&-i\zeta(A_{3}\phi_{1}+A^{-}\phi_{2}),  \label{34} \\
\phi_{2x}&=&-i\zeta(A^{+}\phi_{1}-A_{3}\phi_{2}), \label{35} 
\end{eqnarray}
where
\begin{eqnarray}
F=\frac{2}{\alpha}\left(
\begin{array}{cc}
A^{-}\bar{\phi}_{1}\phi_{2}-A^{+}\phi_{1}\bar{\phi}_{2} & 2A_{3}\phi_{1}\bar{\phi}_{2}-A^{-}(|\phi_{1}|^{2}-|\phi_{2}|^{2})\\
A^{+}(|\phi_{1}|^{2}-|\phi_{2}|^{2})-2A_{3}\bar{\phi}_{1}\phi_{2} & A^{+}\phi_{1}\bar{\phi}_{2}- A^{-}\bar{\phi}_{1}\phi_{2} \end{array}
\right). \label{36} 
\end{eqnarray}

\subsection{Lax representation}
The M-XIV equation (\ref{9})-(\ref{10}) is integrable. Its LR reads as
\begin{eqnarray}
\Phi_{x}&=&U_{2}\Phi,\label{37} \\ 
\Phi_{t}&=&V_{2}\Phi,\label{38}
\end{eqnarray}
where
\begin{eqnarray}
U_{2}&=&-i(\lambda-\alpha)A,\label{39} \\ 
V_{2}&=&-\frac{i(\lambda-\alpha)}{\alpha\lambda}\left(\alpha^{2}fA-\frac{i\alpha}{4}[A,A_{t}]-\frac{\alpha}{\alpha-\omega}W\right)-\frac{i(\lambda-\alpha)}{(\alpha-\omega)(\lambda-\omega)}W.\label{40}
\end{eqnarray}

\subsection{Reduction}

Let $W=0$. Then the M-XIV  equation takes the form
\begin{eqnarray}
iA_{t}&=&i(fA)_{x}+\frac{1}{4\alpha}[A,A_{t}]_{x}\label{41} 
\end{eqnarray}
or
\begin{eqnarray}
iA_{t}&=&ifA_{x}+\frac{1}{4\alpha}[A,A_{tx}], \label{42} \\
f_{x}&=&\frac{1}{4i\alpha}tr(A[A_{t},A_{x}]), \label{43}
\end{eqnarray}
and which is the so-called M-XIII equation \cite{1812.02152}.

\subsection{Conservation laws}
As the integrable system, the M-XIV equation admits the infinity number of conservation laws. The simplest one that  we can get, for example, from Eqs. (\ref{18})-(\ref{20})  has the following form
\begin{eqnarray}
({\bf A}_{x}^{2})_{t}+(8{\bf A}\cdot{\bf W}+8\alpha^{2}f)_{x}=0. \label{44}
\end{eqnarray}
In  ${\bf W}=0$  case, this equation takes the form
\begin{eqnarray}
({\bf A}_{x}^{2})_{t}+8\alpha^{2}f_{x}=0 \label{45}
\end{eqnarray}
so that
\begin{eqnarray}
({\bf A}_{x}^{2})_{t}=4\alpha{\bf A}\cdot ({\bf A}_{x} \wedge {\bf A}_{t}). \label{46}
\end{eqnarray}

\section{Integrable motion of space curves. The equation Lakshmanan (geometrical) equivalent to the M-XIV equation}

In this section, we  establish the  link between the M-XIV  equation   and   the motion of space curves. Then using this link we can find the Lakshmanan (geometrical) equivalent counterpart of the M-XIV  equation. Consider a family of smooth space curve in $R^{3}$ which we define as 
\begin{eqnarray}
{\bf \gamma} (x,t): [0,X] \times [0, T] \rightarrow R^{3},\label{47}
\end{eqnarray} 
where $x$ is the arc length of the curve at each time $t$.  In this case, the unit tangent vector ${\bf e}_{1}$,  principal normal vector ${\bf e}_{2}$ and binormal vector ${\bf e}_{3}$ are defined as
\begin{eqnarray}
{\bf e}_{1}={\bf \gamma}_{x}, \quad {\bf e}_{2}=\frac{{\bf \gamma}_{xx}}{|{\bf \gamma}_{xx}|}, \quad {\bf e}_{3}={\bf e}_{1}\wedge {\bf e}_{2}, \label{48}
\end{eqnarray} 
respectivily. The corresponding  Frenet-Serret equation reads as 
 \begin{eqnarray}
\left ( \begin{array}{ccc}
{\bf  e}_{1} \\
{\bf  e}_{2} \\
{\bf  e}_{3}
\end{array} \right)_{x} = C
\left ( \begin{array}{ccc}
{\bf  e}_{1} \\
{\bf  e}_{2} \\
{\bf  e}_{3}
\end{array} \right)=
\left ( \begin{array}{ccc}
0   & \kappa_{1}     & \kappa_{2} \\
-\kappa_{1}& 0     & \tau  \\
-\kappa_{2}   & -\tau & 0
\end{array} \right)\left ( \begin{array}{ccc}
{\bf  e}_{1} \\
{\bf  e}_{2} \\
{\bf  e}_{3}
\end{array} \right), \label{49} 
\end{eqnarray}
where $\tau$,  $\kappa_{1}$ and $\kappa_{2}$ are   torsion,  geodesic curvature and  normal curvature of the curve, respectively.

It is well-known that between some integrable  systems there take place the  geometrical (Lakshmanan) and gauge equivalences. 
The Frenet-Serret equation and the general temporal evolution equation are given by 
\begin{eqnarray}
\left ( \begin{array}{ccc}
{\bf  e}_{1} \\
{\bf  e}_{2} \\
{\bf  e}_{3}
\end{array} \right)_{x} = C
\left ( \begin{array}{ccc}
{\bf  e}_{1} \\
{\bf  e}_{2} \\
{\bf  e}_{3}
\end{array} \right),\quad
\left ( \begin{array}{ccc}
{\bf  e}_{1} \\
{\bf  e}_{2} \\
{\bf  e}_{3}
\end{array} \right)_{t} = G
\left ( \begin{array}{ccc}
{\bf  e}_{1} \\
{\bf  e}_{2} \\
{\bf  e}_{3}
\end{array} \right), \label{50} 
\end{eqnarray}
where
\begin{eqnarray}
C =
\left ( \begin{array}{ccc}
0   & \kappa_{1}     & \kappa_{2}  \\
-\kappa_{1}  & 0     & \tau  \\
-\kappa_{2}    & -\tau & 0
\end{array} \right) ,\quad
G =
\left ( \begin{array}{ccc}
0       & \omega_{3}  & \omega_{2} \\
-\omega_{3} & 0      & \omega_{1} \\
-\omega_{2}  & -\omega_{1} & 0
\end{array} \right).\label{51} 
\end{eqnarray}
The compatibility condition of these equations is written as
\begin{eqnarray}
C_t - G_x + [C, G] = 0\label{52} 
\end{eqnarray}
or in elements   
 \begin{eqnarray}
\kappa_{1t}- \omega_{3x} -\kappa_{2}\omega_{1}+ \tau \omega_2&=&0, \label{53} \\ 
\kappa_{2t}- \omega_{2x} +\kappa_{1}\omega_{1}- \tau \omega_3&=&0, \label{54} \\
\tau_{t}  -    \omega_{1x} - \kappa_{1}\omega_2+\kappa_{2}\omega_{3}&=&0.  \label{55} \end{eqnarray}
Our next step is the following identification
 \begin{eqnarray}
{\bf A}\equiv {\bf e}_{1} \label{56} 
\end{eqnarray}
and the assumption    
\begin{eqnarray}
\kappa_{1}=-2\zeta, \quad \kappa_{2}=r-q, \quad \tau=-i(r+q), \label{57} 
\end{eqnarray}
where $q=-0.5(\kappa_{2}-i\tau)$ and $r=0.5(\kappa_{2}+i\tau)$ are some  functions,  $\zeta=const.$  Then we have 
\begin{eqnarray}
\omega_{1} & = &\frac{0.5r_{t}-0.5q_{t}-n-p)}{\zeta}+\frac{n+p}{\zeta-\omega},\label{58}\\ 
\omega_{2}&=& \frac{i(0.5r_{t}+0.5q_{t}-n+p)}{\zeta}+\frac{n-p}{\zeta-\omega}, \label{59} \\
\omega_{3} & = &\frac{2\eta}{\zeta-\omega}+\frac{2a}{\zeta}.      \label{60}
\end{eqnarray}
The Eqs.(\ref{33})-(\ref{35}) give us the following equations for $q, r, \rho=4a, n, p, \eta$: \begin{eqnarray}
q_{xt}-\rho q+4p_{x}&=&0, \label{61} \\
r_{xt}-\rho r-4n_{x}&=&0, \label{62} \\
\rho_{x}-0.5(rq)_{t}+2(qn-rp)&=&0,\label{63}\\
\eta_{x}+0.5(rp-qn)&=&0, \label{64} \\
p_{x}+2i\omega p+\eta q&=&0, \label{65}\\
n_{x}-2i\omega n -\eta r&=&0 \label{66}
\end{eqnarray}
which is the so-called Myrzakulov-XXXII (M-XXXII) equation. Note that for the reductions $r=\sigma \bar{q}$ and $n=-\sigma{\bar p}$,  the M-XXXII equation (\ref{44})-(\ref{49}) takes the form
\begin{eqnarray}
q_{xt}-4aq+2p_{x}&=&0, \label{67} \\
a_{x}-\sigma[0.5(|q|^{2})_{t}+q\bar{p}+\bar{q}p]&=&0,\label{68}\\
\eta_{x}+\sigma({\bar q}p+q{\bar p})&=&0, \label{69} \\
p_{x}+2i\omega p+2\eta q&=&0. \label{70}
\end{eqnarray}

So, we have  proved  the  Lakshmanan (geometrical) equivalence between the M-XIV  equation  (\ref{9})-(\ref{10}) and the M-XXXII equation (\ref{67})-(\ref{70}). 

\section{On the equation gauge equivalent to the M-XIV  equation}

Taking into account the  results of the previous section, we expect that  there exists gauge equivalence between the M-XIV equation (\ref{9})-(\ref{10}) and the M-XXXII equation (\ref{44})-(\ref{49}). To check it,  consider the following transformation
$\Phi=g\Psi$, where $\Phi$ is some solutions of the set (\ref{37})-(\ref{38}) and $g=\Psi|_{\lambda=\alpha}$.
Then the function $\Psi$ obeys the following set of linear equations
\begin{eqnarray}
\Psi_{x}&=&U_{5}\Psi,\label{71} \\ 
\Phi_{t}&=&V_{5}\Psi.\label{72}
\end{eqnarray}
Here
\begin{eqnarray}
U_{5}=-i\lambda \sigma_{3}+Q, \quad V_{5}=\frac{i}{\lambda}F+\frac{i}{\lambda-\omega}G, \label{73}
\end{eqnarray}
where
\begin{eqnarray}
Q=\left(
\begin{array}{cc}
0 & q \\
\sigma{\bar q} & 0
\end{array}
\right), \quad F=\left(
\begin{array}{cc}
a & -0.5q_{t}-p \\
0.5\sigma({\bar q}_{t}+{\bar p}) & -a
\end{array}
\right), \quad G=\left(\begin{array}{cc}
\eta & p \\
-\sigma{\bar p} & -\eta
\end{array}
\right). \label{74}
\end{eqnarray}
It can be easily shown that the compatibility condition $U_{5t}-V_{5x}+[U_{5}, V_{5}]=0$ gives   the M-XXXII equation (\ref{61})-(\ref{66}).

\section{The  M-XXXII equation}

For our convenience, here we collect the main formulas of the  M-XXXII equation  (\ref{61})-(\ref{66}). 

\subsection{Equation}
The  M-XXXII equation  (\ref{44})-(\ref{49})  reads as
\begin{eqnarray}
q_{xt}-4aq+2p_{x}&=&0, \label{75} \\
r_{xt}-4ar-2n_{x}&=&0, \label{76} \\
a_{x}-0.5(rq)_{t}+qn-rp&=&0,\label{77}\\
\eta_{x}+rp-qn&=&0, \label{78} \\
p_{x}+2i\omega p+2\eta q&=&0, \label{79}\\
n_{x}-2i\omega n -2\eta r&=&0. \label{80}
\end{eqnarray}

\subsection{Lax representation}
The M-XXXII  equation  is integrable and its LR is 
\begin{eqnarray}
\Psi_{x}&=&U_{3}\Psi,\label{81} \\ 
\Psi_{t}&=&V_{3}\Psi.\label{82}
\end{eqnarray}
Here
\begin{eqnarray}
U_{3}=-i\lambda \sigma_{3}+Q, \quad V_{3}=\frac{i}{\lambda}F+\frac{i}{\lambda-\omega}G, \label{83}
\end{eqnarray}
where
\begin{eqnarray}
Q=\left(
\begin{array}{cc}
0 & q \\
r & 0
\end{array}
\right), \quad F=\left(
\begin{array}{cc}
a & -0.5q_{t}-p \\
0.5r_{t}-n & -a
\end{array}
\right), \quad G=\left(\begin{array}{cc}
\eta & p \\
n & -\eta
\end{array}
\right). \label{84}
\end{eqnarray}
Note that from Eqs. (\ref{78})-(\ref{80}) we get  the following important formula
\begin{eqnarray}
\eta^{2}+np =const \label{85}
\end{eqnarray}
or, for simplicity,
\begin{eqnarray}
\eta^{2}+np =1. \label{86}
\end{eqnarray}

\subsection{Reductions}
\subsubsection{Case 1: $r=\sigma q$}
First we consider the particular case when 
\begin{eqnarray}
r=\sigma q. \label{87}
\end{eqnarray}
Then we have
\begin{eqnarray}
q_{xt}-4aq+2p_{x}&=&0, \label{88} \\
\eta_{x}+2\sigma qp&=&0, \label{89} \\
p_{x}+2i\omega p+2\eta q&=&0, \label{90}
\end{eqnarray}
where $n=-\sigma  p$ and $\omega={\bar \omega}$.
This  equation  is also integrable and its LR is 
\begin{eqnarray}
\Psi_{x}&=&U_{4}\Psi,\label{91} \\ 
\Psi_{t}&=&V_{4}\Psi.\label{92}
\end{eqnarray}
Here
\begin{eqnarray}
U_{4}=-\lambda \sigma_{3}+Q, \quad V_{4}=\frac{i}{\lambda}F+\frac{i}{\lambda-\omega}G, \label{93}
\end{eqnarray}
where
\begin{eqnarray}
Q=\left(
\begin{array}{cc}
0 & q \\
\sigma  q & 0
\end{array}
\right), \quad F=\left(
\begin{array}{cc}
a & -0.5q_{t}-p \\
0.5\sigma  q_{t}+\sigma  p & -a
\end{array}
\right), \quad G=\left(\begin{array}{cc}
\eta & p \\
\sigma  p & -\eta
\end{array}
\right). \label{94}
\end{eqnarray}

\subsubsection{Case 2: $r=\sigma \bar{q}$}
In this case that is when 
\begin{eqnarray}
r=\sigma \bar{q}, \quad n=-\sigma{\bar p}, \label{95}
\end{eqnarray} we get the following set of equations
\begin{eqnarray}
q_{xt}-4aq+2p_{x}&=&0, \label{96} \\
a_{x}-\sigma[0.5(|q|^{2})_{t}+q\bar{p}+\bar{q}p]&=&0,\label{97}\\
\eta_{x}+\sigma({\bar q}p+q{\bar p})&=&0, \label{98} \\
p_{x}+2i\omega p+2\eta q&=&0. \label{99}
\end{eqnarray}
This reduction of the  M-XXXII equation  is integrable with the following  LR 
\begin{eqnarray}
\Psi_{x}&=&U_{5}\Psi,\label{100} \\ 
\Phi_{t}&=&V_{5}\Psi.\label{101}
\end{eqnarray}
Here
\begin{eqnarray}
U_{5}=-\lambda \sigma_{3}+Q, \quad V_{5}=\frac{i}{\lambda}F+\frac{i}{\lambda-\omega}G,
\end{eqnarray} \label{102}
where
\begin{eqnarray}
Q=\left(
\begin{array}{cc}
0 & q \\
\sigma{\bar q} & 0
\end{array}
\right), \quad F=\left(
\begin{array}{cc}
a & -0.5q_{t}-p \\
0.5\sigma({\bar q}_{t}+{\bar p}) & -a
\end{array}
\right), \quad G=\left(\begin{array}{cc}
\eta & p \\
-\sigma{\bar p} & -\eta
\end{array}
\right). \label{103}
\end{eqnarray}

\subsubsection{Case 3: $p=n=\eta=0$}
In this  particular case when 
\begin{eqnarray}
p=n=\eta=0,  \label{104}
\end{eqnarray}
 we have
\begin{eqnarray}
q_{xt}-4aq&=&0, \label{105} \\
r_{xt}-4ar&=&0, \label{106} \\
a_{x}-0.5(rq)_{t}&=&0,\label{107}
\end{eqnarray}
which is, in fact, the GKOE (\ref{3})-(\ref{5}).

\subsection{The M-XXXII equation as the equation with self-consistent sources}
We can rewrite the M-XXXII equation as the equation with self-consistent sources. Let us consider the following representation for the functions $p$, $n$ and $\eta$:
\begin{eqnarray}
p=2\psi_{1}\bar{\psi}_{2}, \quad n=-2\sigma\bar{\psi}_{1}\psi_{2}, \quad \eta=|\psi_{1}|^{2}+\sigma|\psi_{2}|^{2}, \label{108}
\end{eqnarray}
where $\psi_{1}|^{2}-\sigma|\psi_{2}|^{2}=const$ and $\Psi=(\psi_{1}, \psi_{2})^{T}$ satisfy the equations (\ref{81})-(\ref{82}). Then the M-XXXII equation takes the form
\begin{eqnarray}
q_{xt}-4aq+4(\psi_{1}\bar{\psi}_{2})_{x}&=&0, \label{109} \\
a_{x}-0.5\sigma(|q|^{2})_{t}-2\sigma(q\bar{\psi}_{1}\psi_{2}+\bar{q}\psi_{1}\bar{\psi}_{2})&=&0,\label{110}\\
\psi_{1x}+i\xi \psi_{1}-q\psi_{2}&=&0, \label{111} \\
\psi_{2x}-i\xi \psi_{2}-\sigma \bar{q}\psi_{1}&=&0 \label{112}
\end{eqnarray}
or
\begin{eqnarray}
q_{xt}-4aq-8i\xi\psi_{1}\bar{\psi}_{2}+4q(\sigma|\psi_{1}|^{2}+|\psi_{2}|^{2})&=&0, \label{113} \\
a_{x}-0.5\sigma(|q|^{2})_{t}-2\sigma(q\bar{\psi}_{1}\psi_{2}+\bar{q}\psi_{1}\bar{\psi}_{2})&=&0,\label{114}\\
\psi_{1x}+i\xi \psi_{1}-q\psi_{2}&=&0, \label{115} \\
\psi_{2x}-i\xi \psi_{2}-\sigma \bar{q}\psi_{1}&=&0. \label{116}
\end{eqnarray}
This is the desired form of the M-XXXII equation written as the equation with self-consistent sources.
\section{Scale transformation}
Consider the following scale transformation
\begin{eqnarray}
a\rightarrow 0.25 \rho, \quad (q,r)\rightarrow 0.5(q,r).\label{117}
\end{eqnarray}
Under this transformation, the  M-XXXII equation (\ref{44})-(\ref{49}) becomes
\begin{eqnarray}
q_{xt}-\rho q+4p_{x}&=&0, \label{118} \\
r_{xt}-\rho r-4n_{x}&=&0, \label{119} \\
\rho_{x}-0.5(rq)_{t}+2(qn-rp)&=&0,\label{120}\\
\eta_{x}+0.5(rp-qn)&=&0, \label{121} \\
p_{x}+2i\omega p+\eta q&=&0, \label{122}\\
n_{x}-2i\omega n -\eta r&=&0. \label{123}
\end{eqnarray}
When $p=n=\eta=0$, it looks like
\begin{eqnarray}
q_{xt}-\rho q&=&0, \label{124} \\
r_{xt}-\rho r&=&0, \label{125} \\
\rho_{x}-0.5(rq)_{t}&=&0,\label{126}
\end{eqnarray}
which is a more standard form of the GKOE (\ref{3})-(\ref{5}) with $(r)\rightarrow (-r)$.

\section{Integrable surface induced by the M-XIV  equation}

In this section, we  present a surface which is induced by the M-XIV equation. Let us consider  a surface in three-dimensional Euclidean space $R^{3}$ that is parametrized by position vector ${\bf r}(x,t)$ of the surface. To construct the surface, let us   do  the following identification
\begin{eqnarray}
{\bf A}\equiv{\bf r}_{x}, \label{127} 
\end{eqnarray}
where ${\bf r}_{x}^{2}=1$. After that from the vector M-XIV equation (\ref{16})-(\ref{17}) we get the following Myrzakulov-XXXI (M-XXXI) equation
\begin{eqnarray}
{\bf r}_{t}&=&f{\bf r}_{x}+\frac{1}{2\alpha}{\bf r}_{x} \wedge {\bf r}_{xt}+\frac{1}{\alpha(\omega-\alpha)} {\bf W}, \label{128}\\
f_{x}&=&\frac{1}{2\alpha}{\bf r}_{x}\cdot ({\bf r}_{xt} \wedge {\bf r}_{xx}), \label{129}\\
{\bf W}_{x}&=&2(\omega-\alpha){\bf r}_{x} \wedge {\bf W}, \label{130}
\end{eqnarray}
where ${\bf r}=(r_{1}, r_{2}, r_{3})$.
The M-XXXI equation (\ref{128})-(\ref{130}) is integrable. Its LR looks as 
\begin{eqnarray}
\Phi_{x}&=&U_{7}\Phi,\label{131} \\ 
\Phi_{t}&=&V_{7}\Phi,\label{132}
\end{eqnarray}
where
\begin{eqnarray}
U_{7}&=&-i(\lambda-\alpha)r_{x},\label{133} \\ 
V_{7}&=&-\frac{i(\lambda-\alpha)}{\alpha\lambda}\left(\alpha^{2}fr_{x}-\frac{i\alpha}{4}[r_{x},r_{xt}]-\frac{\alpha}{\alpha-\omega}W\right)-\frac{i(\lambda-\alpha)}{(\alpha-\omega)(\lambda-\omega)}W.\label{134}
\end{eqnarray}
The compatibility condition of the equations (\ref{131})-(\ref{132}) gives  the following equation 
\begin{eqnarray}
ir_{t}&=&ifr_{x}+\frac{1}{4\alpha}[r,r_{t}]+\frac{i}{\alpha(\alpha-\omega)}W, \label{135} \\
f_{x}&=&\frac{1}{4i\alpha}tr(r_{x}[r_{xt},r_{xx}]), \label{136} \\
W_{x}&=&i(\alpha-\omega)[r_{x},W] \label{137} 
\end{eqnarray}
which is the matrix form of the M-XXXI equation.
Here 
\begin{eqnarray}
r&=&\left(
\begin{array}{cc}
r_{3} &  r^{-} \\
r^{+} & -r_{3}
\end{array}
\right), \quad r_{x}^{2}=I, \quad r_{1x}^{2}+r_{2x}^{2}+r_{3x}^{2}=1, \quad r^{\pm}=r_{1}\pm i r_{2}.  \label{138}
\end{eqnarray}
In our case, the first fundamental form of the surface is given by
\begin{eqnarray}
I=dx^{2}+2({\bf r}_{x}\cdot {\bf r}_{t})dxdt+{\bf r}_{t}^{2}dt^{2}, \label{139}
\end{eqnarray}
where 
\begin{eqnarray}
{\bf r}_{x}^{2}&=&1, \label{140}\\
{\bf r}_{x}\cdot {\bf r}_{t}&=&f+\frac{1}{\alpha(\omega-\alpha)} {\bf r}_{x}\cdot{\bf W}, \label{141}\\
{\bf r}_{t}^{2}&=&f^{2}+\frac{1}{\alpha^{2}(\omega-\alpha)^{2}}+\frac{2f({\bf r}_{x}\cdot{\bf W})}{\alpha(\omega-\alpha)} +\frac{|{\bf r}_{x} \wedge {\bf r}_{xt}|^{2}}{4\alpha^{2}}+\frac{({\bf r}_{x}\wedge{\bf r}_{xt})\cdot{\bf W}}{\alpha^{2}(\omega-\alpha)}. \label{142}
\end{eqnarray}
Finally, we note that the M-XXXI equation (\ref{128})-(\ref{130}) can be rewritten as
\begin{eqnarray}
{\bf r}_{xt}&=&2\alpha{\bf r}_{t} \wedge {\bf r}_{x}+\frac{2}{\alpha-\omega} {\bf r}_{x}\wedge{\bf W}, \label{143}\\
{\bf W}_{x}&=&2(\omega-\alpha){\bf r}_{x} \wedge {\bf W}, \label{144}
\end{eqnarray}
or
\begin{eqnarray}
{\bf r}_{xt}&=&2\alpha{\bf r}_{t} \wedge {\bf r}_{x}-\frac{1}{(\alpha-\omega)^{2}} {\bf W}_{x}, \label{145}\\
{\bf W}_{x}&=&2(\omega-\alpha){\bf r}_{x} \wedge {\bf W}. \label{146}
\end{eqnarray}
The above obtained results are enough to define the surface in  three-dimensional Euclidean space $R^{3}$  parametrized by the position vector ${\bf r}(x,t)$ of the surface. This surface is integrable as its equation that is the equation for ${\bf r}$ (\ref{128})-(\ref{130}) admits LR. So we have shown that the  M-XXXI equation induces the some integrable surface. Lastly, we present the reduction of the M-XXXI equation. If  ${\bf W}=0$, then it turns to the following equation  \cite{1603.00781}
\begin{eqnarray}
{\bf r}_{t}=f{\bf r}_{x}+\frac{1}{2\alpha}{\bf r}_{x} \wedge {\bf r}_{xt} \label{147}
\end{eqnarray}
or
\begin{eqnarray}
{\bf r}_{xt}=2\alpha{\bf r}_{t} \wedge {\bf r}_{x}, \label{148}
\end{eqnarray}
where 
\begin{eqnarray}
f={\bf r}_{t}\cdot {\bf r}_{x}. \label{149}
\end{eqnarray}
Finally,  let us  present the M-XXXI equation (\ref{128})-(\ref{130}) as the equation with self-consistent sources.  It has the form
\begin{eqnarray}
 r_{1t}&=&f r_{1x}+\frac{1}{2\alpha}(r_{2x}r_{3xt}-r_{2xt}r_{3x})+\frac{1}{\alpha(\omega-\alpha)}(\phi_{1}\bar{\phi}_{2}+\bar{\phi}_{1}\phi_{2}), \label{150}\\
r_{2t}&=&f r_{2x}+\frac{1}{2\alpha}(r_{3x}r_{1xt}-r_{3xt}r_{1x})+\frac{i}{\alpha(\omega-\alpha)}(\phi_{1}\bar{\phi}_{2}-\bar{\phi}_{1}\phi_{2}), \label{151}\\
r_{3t}&=&f r_{3x}+\frac{1}{2\alpha}(r_{1x}r_{2xt}-r_{1xt}r_{2x})+\frac{1}{\alpha(\omega-\alpha)}(|\phi_{1}|^{2}-|\phi_{2}|^{2}
), \label{152} \\
f_{x}&=&\frac{1}{4i\alpha}tr(r_{x}[r_{xt},r_{xx}]), \label{153} \\
\phi_{1x}&=&-i\zeta(r_{3x}\phi_{1}+r^{-}_{x}\phi_{2}), \label{154}  \\
\phi_{2x}&=&-i\zeta(r_{x}^{+}\phi_{1}-r_{3x}\phi_{2}). \label{155} 
\end{eqnarray}
It is the M-XXXI equation written in the form of the equation with self-consistent sources. 
\section{Conclusions}
In this paper, we have shown that the HFE  equation admits an integrable  generalization with the self-consistent sources - the M-XIV equation.  In particular,
the  integrability of the M-XIV equation  has been established by constructing its Lax pair. We have also demonstrated that the M-XIV  equation is equivalent  to the M-XXXII equation. At the level,  some reductions  of the  M-XXXII equation    are found.  Another interesting issue is to generalize these results  to the real and complex short pulse equations \cite{R2019}.  
 Although in this paper, we have restricted our consideration to the mathematical aspects of the proposed integrable two equations, 
the relevance of these equations  as  models capable of describing the dynamics of ultra-short pulses in optical fibers  is an important issue to be studied in a future work.

\section{Acknowledgements}
This work was supported  by  the Ministry of Edication  and Science of Kazakhstan under
grants 0118РК00935 and 0118РК00693.

\end{document}